\documentclass{PoS}

\usepackage{multirow}

\usepackage{amsmath, amsthm, amssymb}

\usepackage{epsfig}
\usepackage{cite}

\newcommand{\be}{\begin{eqnarray}}
\newcommand{\ee}{\end{eqnarray}}
\newcommand{\bi}{\begin{itemize}}
\newcommand{\ei}{\end{itemize}}

\def\<{\left\langle}
\def\>{\right\rangle}

\def\ls{\left[}
\def\rs{\right]}

\def\ls{\left[}
\def\rs{\right]}

\title{ Features and implications of the plateau inflationary potentials }

\ShortTitle{Features and implications of the plateau inflationary potentials}

\author{\speaker{Ioannis Dalianis}\\

        Physics Division, National Technical University of Athens, \\ 15780 Zografou Campus, Athens, Greece\\
        E-mail: \email{dalianis@mail.ntua.gr}}

\abstract{ After the last P{\scshape lanck} CMB data the plateau inflationary potentials are
favored. I give some examples of such inflationary models emphasizing particularly on the Starobinsky model and its supergravity embedding. I discuss the crucial implications, regarding the initial conditions problem,
of this new sort of potentials for the standard picture of the inflationary theory.

}

\FullConference{ 18th International Conference From the Planck Scale to the Electroweak Scale \\
25-29 May 2015\\
Ioannina, Greece
}

\begin{document}

\section{Introduction to the initial conditions problem}

The new results obtained by P{\scshape lanck} collaboration \cite{Ade:2013uln, Ade:2015lrj} have favored a new class of inflationary models characterized by a special feature: the part of the potential that implements inflation is a plateau. This class of models includes the Starobinsky model \cite{Starobinsky:1980te}, the Higgs inflation \cite{Bezrukov:2007ep} as well as the so called $\alpha$-attractors \cite{Galante:2014ifa, Kallosh:2015lwa}. On the contrary, many of the inflationary  models that were commonly used in the past have been ruled out with the most notable example  the $V(\phi)=m^2 \phi^2$ large field model which is at the edge of the $99\%$ CL contours of the 2015  P{\scshape lanck} analysis.

The selection of the plateau-like inflationary models and the exclusion of the steep ones 
by the last observational data  is certainly a success of the inflationary theory for it certifies its predictive power. At the  same time, however, the  plateau potentials may  question  the generality of the cosmic inflation phase because special initial conditions seem to be required \cite{Ijjas:2013vea, Guth:2013sya, Linde:2014nna, Ijjas:2014nta,  Mukhanov:2014uwa, Goldwirth:1991rj}.
A representative example of the plateau potentials is the Starobinsky model which is an $f(R)=R+R^2$  gravity theory. In the dual picture it yields a potential for the scalaron $\varphi$ that reads
\begin{equation}\label{starp}
V_{R^2}(\varphi) =V_\text{INF} \left(1- e^{-\sqrt{\frac{2}{3}}\varphi/M_P}\right)^2\,,
\end{equation}
where $V_\text{INF}$  the upper bound of the inflationary energy density,  $V_\text{INF} \sim 10^{-10}M^4_{P} \ll M^4_{P}$.
These potentials are not capable to drive an inflationary phase right after the Planck era implying that our Universe started with a decelerating phase ($\ddot a <0$) instead of an accelerating one ($\ddot a>0$).

It is very motivated the inflationary phase  to initiate close to the Planck energy scale because it assures the natural creation of our observable Universe without rather special initial conditions.  Indeed, even a fundamentally small initial patch of Planck length radius $ l_P$
when dominated by the potential energy of the inflaton field, $\frac12 \dot{\phi}^2 +\frac12(\nabla \phi/a)^2\lesssim V(\phi)\sim \rho_\text{tot}\sim M_P^4$,  starts expanding in an accelerating manner.
The essential implication of this accelerated expansion is the presence of a nearly constant event horizon distance with size  $\sim l_P$, that is, of the order of the curvature scale, the so-called Hubble radius. 
The event horizon {\it protects} the initial smooth patch from the outside inhomogeneous regions where the gradients of the field  are nonzero. Otherwise, if the event horizon had been unbounded,  the inhomogeneities would have propagated and infested the initial smooth patch, spoiling the inflationary phase. 

If inflation is unable to start at energies close to the Planck scale then the minimum size of the initial homogeneous patch has to be much larger than $l_P$. Given the smallness of the  plateau energy density, $V_\text{INF}$, one  has to assume that a kinetic-energy domination regime preceded the inflationary phase.
One finds for plateau potentials that the density inhomogeneities have to be expelled at least $10^3$ Hubble scales farther if the Universe has emerged from the Planck density, $M^4_P \equiv (2.4 \times 10^{18} \text{GeV})^4$ \cite{Dalianis:2015fpa}.

The realization of inflation requires a homogeneous patch of minimum radius about $ H^{-1}(t=t_\text{INF})$
 which can exist only if the primary patch at $t_\text{init}$, $ H^{-1}(t_\text{INF})\,a(t_\text{init})/a(t_{INF})$, is surrounded by a {\it supplementary} homogeneous shell of width equal to the event horizon distance $d_\text{event} (t_\text{init}, t_\text{INF})$ \cite{Goldwirth:1991rj, Kung:1989xz, Vachaspati:1998dy, Dalianis:2015fpa},  see figure 1.
\begin{figure} 
\centering
\includegraphics [scale=.85, angle=0]{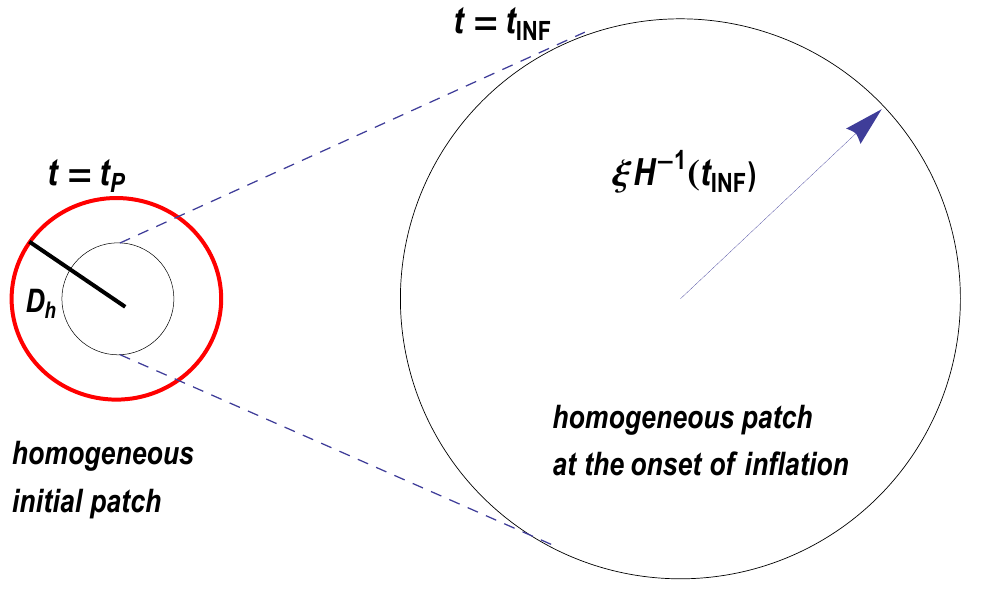} 
\caption{\small{A schematic illustration of the minimum homogeneous patch at two different times: right after the Planck time, $t=t_P$, and at the onset of inflation, $t_\text{INF} \sim 10^{5} t_P$. Its radius at $t_P$ is $D_h\equiv D_\text{homog}(t_P)$ and at $t_\text{INF}$ is $\xi H^{-1}_\text{INF}\sim 10^5 l_P$.}}
\end{figure}

\begin{figure} 
\centering
\includegraphics [scale=.85, angle=0]{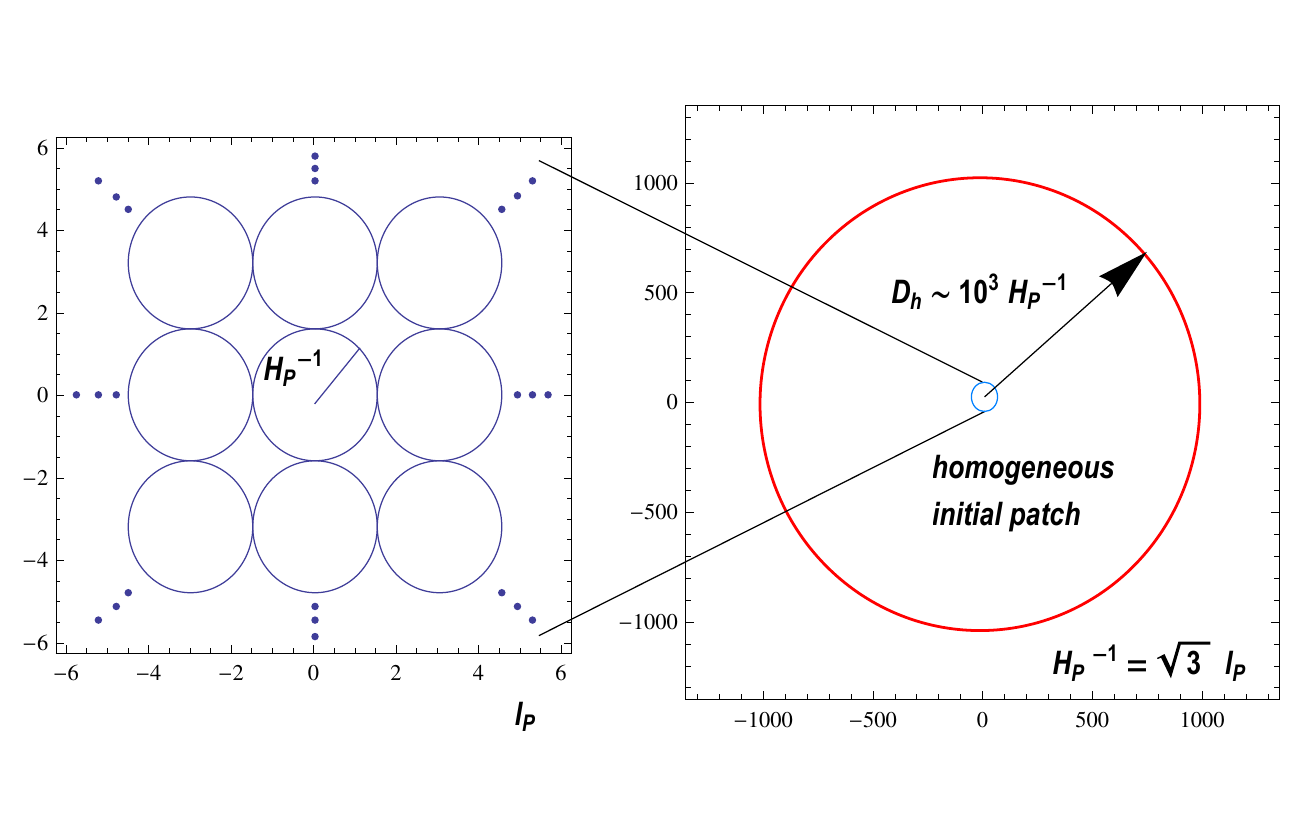} 
\caption{\small{The figure illustrates schematically the initial conditions problem for the plateau inflationary models as the Starobinsky $R^2$ where $V_\text{INF} \sim 10^{-10}M^4_
P$. The delayed inflationary dynamics imply that right after the Planck time hundreds of billions of causally disconnected regions (CDR) have to be homogeneous in order not to spoil the onset of inflation ($D_h\equiv D_\text{homog}(t_P)$).}}
\end{figure}

For flat space and $t_\text{init}\sim t_P$ and $t_\text{INF} \sim V^{1/4}_\text{INF}$ the required homogeneous volume, $(4/3)\pi\, D^3_\text{homog}(t_P)$, is at least $10^{11}$ times bigger than $(4/3)\pi l_P^{3}$ which means that, initially,  hundreds of billions of causally disconnected regions were much similar without any dynamical reason. Briefly we call them  {\it CDR (Causally Disconnected Regions)}. We consider the $l_P$ as the {\it causal horizon} at Planck times. We find  that for an open Universe the number of CDR required to be homogeneous is larger compared to the flat case, while for a closed Universe the number is decreased about an order of magnitude, albeit the CDR remains formidable large \cite{Dalianis:2015fpa}. In fact, these are much special initial conditions for the $R^2$ model and any similar plateau potential inflationary models.

There are suggestions for non-trivial pre-inflationary topologies which can address the initial conditions problem, see \cite{Starkman:1998ed, Linde:2004nz, Carrasco:2015rva} for compact flat or open Universe with size which can be less than $H^{-1}$. 
Also, it is possible to design scalar potentials which exhibit different inflationary stages starting  from Planck densities and ending up at a plateau-like part, see e.g. \cite{Artymowski:2015mva}; or to invoke the well motivated non-minimal couplings that yield an effective potential which is flat enough to fit the data with inflation starting from Planck densities \cite{Germani:2010gm,Dalianis:2014nwa, Dalianis:2014sqa, Dalianis:2015aba}. 
Here we  focus on the plateau potentials that are particularly favored by the data and trigger intriguing questions concerning the initial conditions problem. We note that the plateau inflationary trajectory, if sufficiently stable,  can yield $N\gtrsim 60$ e-foldings and no special  fine tuning of the initial values for the fields is required as happens to other models, see e.g. \cite{Tetradis:1997kp, Mendes:2000sq}.

\section{Inflationary models with plateau potentials}

A very motivated approach to the description of inflation is to consider it as a pure gravitational effect.
The Einstein-Hilbert action $S_\text{EH} ={M^2_\text{Pl}}/2 \int d^4x \sqrt{-g}\, R$
 is non-renormalizable and should be understood  as an effective theory that describes the low energy degree of freedom of gravity which is the spacetime metric $g_{\mu\nu}$. Given our ignorance about the UV completion of gravity, the simplest 
extension beyond minimal GR is to write the low-energy effective action including all the terms consistent with general coordinate invariance expanded as
\begin{equation} \label{bEH}
S_\text{eff}=  \int d^4x \sqrt{-g}\left[ M^4_\Lambda + \frac{M^2_\text{Pl}}{2} R+ \alpha R^2 + \beta R^{\mu\nu}R_{\mu\nu}+ \frac{1}{M^2}\left(c R^3+...\right)+...\right]\,.
\end{equation}
If this expansion has any relevance with the inflationary dynamics then the cosmological observations indicate that the $\alpha R^2$ term has to dominate over the other terms. Keeping the $R+ R^2$ terms and neglecting the higher order terms then the Starobinsky model is obtained. This precise cut-off at the quadratic order, although it is fully consistent with the state-of-the-art cosmological data sets, still lacks a concrete theoretical explanation. The addition of any higher order term is enough to ruin plateau inflation \cite{Farakos:2013cqa, Kamada:2014gma}. Still the fact that the $R + R^2$ successfully fits
P{\scshape lanck} results may provide an insight into the form of the action (\ref{bEH}) and the effective description of the fundamental
theory of gravity at the particular energy scales. Moreover, the $R+R^2$ gravity  can be elegantly embedded in a supergravitational framework and accommodate an inflationary phase  \cite{Cecotti:1987sa,Kallosh:2013lkr,Farakos:2013cqa,Ferrara:2013wka, Ferrara:2014yna, Dalianis:2014aya}.

Another type of plateau potentials is the Higgs inflationary model. After the discovery of the first, and only till today, elementary scalar particle at CERN, it is reasonable to postulate that the inflaton field is the Higgs field itself. The Higgs field $\phi$ can implement inflation if it is  non-minimally coupled to gravity with $f(\phi)R$ particularly designed, $f(\phi)=M^2_\text{Pl}+\xi\phi^2$\cite{Bezrukov:2007ep}.  In the Einstein frame the canonically normalized Higgs inflaton experiences a potential identical to the $R^2$ plateau potential. This is actually one of the simplest inflaton models compatible with our knowledge of particle physics. The model is phenomenologically strongly motivated, however, any profound theoretical reasoning for the absence of the higher order terms is eluding.

Also, several broad classes of inflationary potentials that feature a plateau and can fit the P{\scshape lanck} data  have been constructed in the context of supergravity and superconformal theory. The most general class is called $\alpha$-attractors \cite{Galante:2014ifa, Kallosh:2015lwa}. For particular values of the parameter $\alpha$ these models give the same predictions with the Starobinsky and the Higgs inflation model.
The parameter $\alpha$ is related to the inverse curvature of the K\"ahler manifold.

We now turn to  the initial conditions problem focusing on the characteristic plateau potential of the $R + R^2$ pure gravity and pure supergravity models. Our calculations are performed in the Einstein frame, see \cite{Gorbunov:2014ewa} for a different approach.

\section{Initial conditions in $R+R^2$ supergravity}

Minimal supergravity has two different formulations: the old-minimal and the new minimal. One of the objectives is to examine whether the embedding of the Starobinsky model
in minimal supergravity renders it more motivated in terms of the inflationary initial conditions problem. We report an affirmative answer to this question: the initial
conditions are significantly relaxed, however, not fully addressed. The reason is the presence of the
dynamical pure supergravitational auxiliary fields.

The \textbf{old-minimal} supergravity multiplet contains the graviton ($e_m^a$), the gravitino ($\psi_m^\alpha$), 
and a pair of auxiliary fields: the complex scalar $M$ and the real vector $b_m$. 
Supersymmetric Lagrangians with curvature higher derivatives also introduce 
kinematic terms for the ``auxiliary'' fields $M$ and $b_m$.   
The embedding of the Starobinsky model of inflation in old-minimal supergravity 
in a superspace  approach 
consists of reproducing the Lagrangian 
\be
e^{-1} {\cal L} = -\frac{M_P^2}{2} R+ \frac{M_P^2}{12 m^2} R^2   . 
\ee
This is achieved by \cite{Cecotti:1987sa,Kallosh:2013lkr,Farakos:2013cqa,Ferrara:2013wka,Dalianis:2014aya} 
\be
\label{OM}
{\cal L} = -3 M_P^2 \int d^4 \theta \,  E \,   
\ls 1 -  \frac{4}{m^2} {\cal R} \bar {\cal R}+  \frac{\zeta}{3 m^4} {\cal R}^2 \bar {\cal R}^2  \rs . 
\ee
Modifications and further properties can be found in  
\cite{Ellis:2013xoa,Ellis:2013nxa,Ellis:2014gxa,Turzynski:2014tza, Kamada:2014gma,Ketov:2014qha, Ferrara:2014yna, Ketov:2014hya,Terada:2014uia,
Alexandre:2013nqa,Alexandre:2014lla}. 
The superspace Lagrangian has a classically equivalent description as standard supergravity coupled to additional superfields $T$ and $S$. The field $S$ remains strongly stabilized and does not affect the evolution. The real and imaginary parts of $T$ field are related to the $\varphi$ and $b$ that experience the potential
\be \label{oldsugpot}
V_{\text{sugra} R^2}(\varphi, b)= \frac34 m^2 M_P^2 \left( 1  -  e^{- \sqrt \frac23 \varphi/M_P}  \right)^2  
+ \frac12 m^2  e^{-2 \sqrt \frac23 \varphi/M_P}  b^2. 
\ee
The dynamical evolution of the pre-inflationary stage consists of two phases: 
\begin{enumerate}
\item From  $V_{\text{sugra}R^2} \simeq M_P^4 $ to $ V_{\text{sugra} R^2} \gtrsim m^2 M_P^2 $, 
both $\varphi$ and $b$ participate in the evolution with the $\varphi$ field rolling first.
\item  At  $ V_{\text{sugra} R^2} \simeq m^2 M_P^2 $ starts the standard Starobinsky inflationary phase with $\varphi$ driving inflation, 
and $b$ now strongly stabilized and  integrated out. 
\end{enumerate}

In the non-supersymmetric Starobinsky,  the energy density $\rho$ has to be dominated by the kinetic term $\dot{\varphi}^2/2$ as long as  $\rho>V_\text{INF}$ which translates into an equation of state $w=1$.
On the other hand, in the Starobinsky supergravity model the pre-inflation expansion of space is much faster and so the event horizon is much smaller, see figure 3.  We find that the minimum number of the CDR contained in the initial homogeneous region of radius $D_\text{homog}$ is \cite{Dalianis:2015fpa}
\be
\frac{V_\text{flat}(D_\text{homog}, w_{\text{sugra}R^2})}{V_\text{flat}(l_P)} = \frac{\frac43 \pi D^3_\text{homog}}{\frac43 \pi l^3_P} \,  \sim \,  10^{6} \,\, \text{CDR}\,.
\ee
Compared to the non-supersymmetric case, in the $R+R^2$ supergravity the required initial homogeneous volume is about a million times smaller.

\begin{figure} 
\textbf{Evolution of the equation of state  \,\,\,\,\,\,\,\,\,\,\,\,\,\,\,\,\,\,\,\;\;\;\;\;\;\quad\quad Evolution of the scale factor}
\centering
\begin{tabular}{cc}
{(a)} \includegraphics [scale=.65, angle=0]{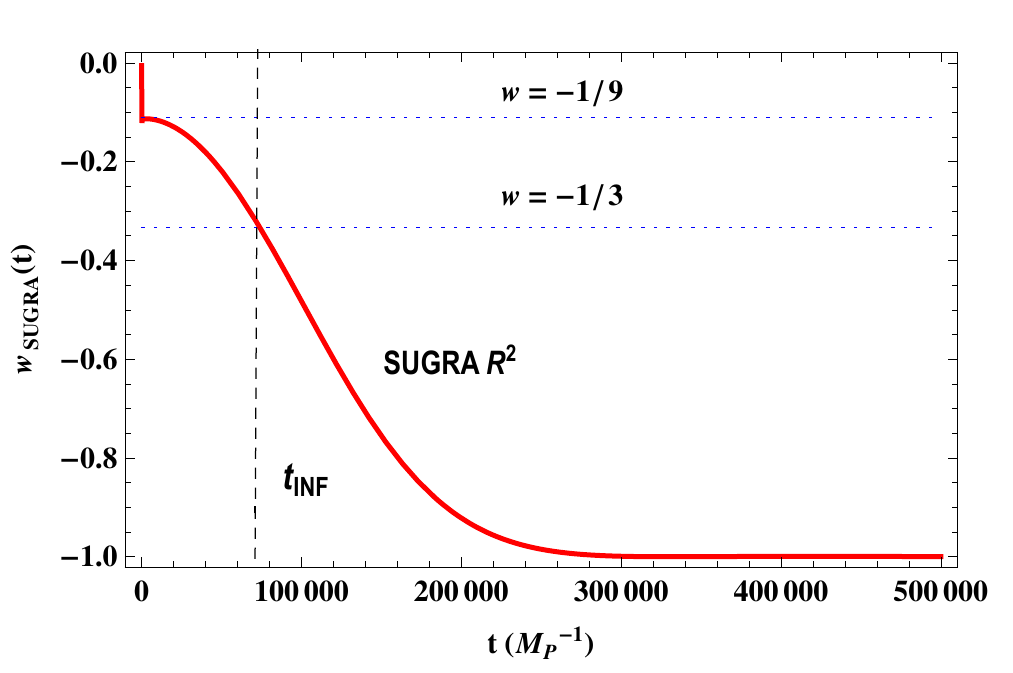} &
{(b)} \includegraphics [scale=.65, angle=0]{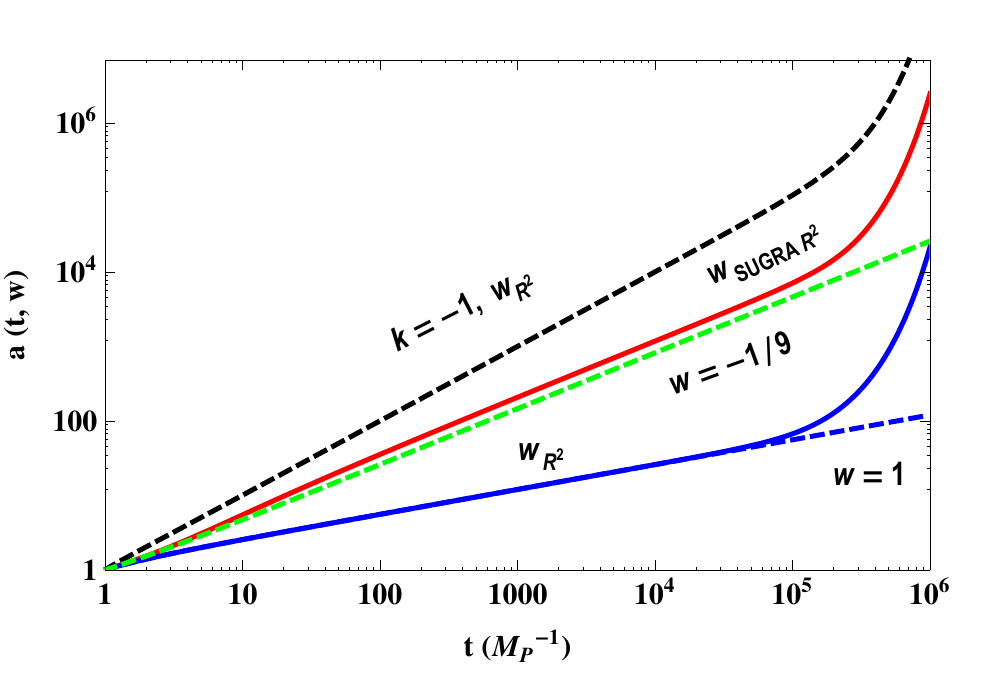}  \\
\end{tabular}
\caption{\small{ {\it The left panel} shows the equation of state, $w$, for the supergravitational system of fields. The initial conditions chosen are equipartition of energy between kinetic and potential thus $w=0$ initially. For some period it is $w\sim -1/9$ and at $t_\text{INF}\simeq 0.7 \times 10^5\, t_P$ the equation of state becomes $w\simeq -1/3$ and acceleration starts. The nearly de-Sitter phase $w\simeq -1 $ starts after $3 \times 10^5 t_P$.  {\it The right panel} shows the evolutions of the cosmological scale factor. The solid lines correspond to solutions for the scale factor of the conventional Starobinsky (lower, blue) and the Starobinsky supergravity (upper, red). The initiation of the accelerating phase is apparent after $t_\text{INF}$. The dashed lines close to the solid ones are the constant equation of state approximations. The lower blue dashed corresponds to the $w=1$ equation of state and describes exactly the evolution of the scale factor before inflation for the $V_{R^2}$ case; the green dashed corresponds to constant $w=-1/9$ which approximates well the $V_{\text{sugra}R^2}$ case until, roughly, the onset of inflation.  The upper black dashed line corresponds to background spatial geometry of negative curvature for the conventional Starobinsky plateau inflationary potential.}}
\end{figure}

The \textbf{new-minimal} supergravity multiplet  contains the graviton field $e_m^a$, 
the gravitino $\psi_m^\alpha$ which are physical fields, a real auxiliary vector $A_m$ which gauges the $U(1)$ R-symmetry and a 
two-form auxiliary field $B_{mn}$. 
The evolution appears similar to the old-minimal case however, here, there is a background vector field with non-vanishing value ${\cal V}_i = {\cal A}_z(t) \delta_{i}^{z}$ which breaks the isotropy of the space. Hence, the scale factor and the event horizon distances in the directions parallel and perpendicular to the vector evolve differently. 

The minimum homogeneous region required at $t_\text{init}=t_P$ for inflation to start at $t_\text{INF}$ in the new-minimal supergravity has volume \cite{Dalianis:2015fpa}
\be \label{oblate}
\begin{split}
V_\text{homog}(t_P) \, =\, 
& \frac43 \pi\, \left(d^{\,xy}_\text{event} (t_P, t_\text{max}) +  \, H^{-1}(t_\text{INF})\frac{a(t_P)}{a(t_{INF})} \right)^2  \\
& \times \, \left(d^{\,z}_\text{event} (t_P, t_\text{max})  + \, H^{-1}(t_\text{INF})\frac{c(t_P)}{c(t_{INF})}  \right) 
\end{split}
\ee
where $d^{xy}_\text{event}$ the event horizon in the  $x-y$ plane, $d^{z}_\text{event}$ in the $z$-direction and $a$, $c$ the two scale factors. At the onset of inflation the contribution of the vector fields in the energy density is subdominant and the initial patch that gets inflated has a spherical volume $V=\frac43 \pi H^{-3}(t_\text{INF})$. However, the initial homogeneous volume (\ref{oblate}) is an oblate spheroid. The number of the causally disconnected regions is found numerically to be of the order 
\be
\frac{V_\text{flat, homog}(t_P)}{\frac43 \pi l_P^3} \, \sim \, 10^7 \,\, \text{CDR}\,.
\ee

\subsection{The curvature term}
The present data find no evidence for any departure from a spatially flat geometry \cite{Planck:2015xua}.
 It is actually inflation itself that addresses the puzzle of the observed flatness of the Universe. Before inflation a homogeneous initial patch is expected to feature either a closed or an open FLRW geometry.
\\ 
\\
\textbf{Closed Universe} \\
When the Universe has a positively curved geometry there is a moment $t_\text{turn}$ the Universe reaches its maximum size and the evolution turns from  expansion to collapse. Inflation has to start before $t_\text{turn}$  that is $\rho(t_\text{INF})>\rho_\text{closed}$ and we find the initial radius.  In the Starobinsky model
 when the initial Universe emerged from the Planck, possibly quantum gravity, era it must have had a radius $a(t_P)= a_\text{init}$ of at least a few thousand times the fundamental Planck length. This is a formidable radius for theories that attempt a quantum description of the genesis of our Universe \cite{Linde:1983cm, Vilenkin:1984wp, Linde:2007fr}.
If it had any smaller size it would have collapsed before inflation begins.

In the Starobinsky supergravity model, potential energy density values $V_{\text{sugra} R^2}(\varphi) \gg V_\text{INF}$ are possible. The lower bound on the minimum radius of the 3-sphere is found to be about 17 times the Planck length. 
The entire volume $V^\text{tot}_\text{closed}$ contains  
\be
\frac{V^\text{tot}_\text{closed}}{V_\text{closed}(l_P)} \simeq \frac{ 2\pi^2 a^3_\text{init,min}}{\frac43 \pi l^3_P }\sim 2\times 10^4\,\, \text{CDR} \,,
\ee
for $t_\text{init}=t_P$
which is again millions of times less severe condition.
\\
\\
\textbf{Open Universe}
\\
When the Universe is open  the space corresponds to a hyperbolic plane that has an infinite volume. We mention that here we are interested only in the local geometry of space inside the Hubble radius not globally. The imaginary radius of the open Universe can take small values such as the Planck length and expand fast and endlessly. The curvature term can dominate over the energy density and dilute the matter till the energy is redshifted to the value of the inflationary plateau $V_\text{INF}$. Then the inflationary evolution takes over, spacetime becomes approximately de Sitter and the negative curvature term asymptotically vanishes. However, if the initial radius of curvature $a(t_P)$ is about the Planck length then the volume enclosed in a sphere of radius $D_\text{homog}\sim 14\, a(t_P)$ is remarkably large in such a highly curved hyperbolic space. 
We find that the number of the causally disconnected regions is astonishingly large
\be \label{opCDR}
\frac{V_\text{open}(D_\text{homog})}{V_\text{open}(l_P)} \sim 2.7 \times 10^{12} \,\,\text{CDR}\,, 
\ee
at the initial time $t_\text{init}=t_P$.

\begin{figure} 
\textbf{\,\,\,\,\,\,\,\,\,\,\,\,\,\,\,\,\,\,\, Number of the causally disconnected regions (\# CDR) \,\,\,\,\;\;\;\;\;\;\;\quad \quad}
\centering
\includegraphics [scale=.75, angle=0]{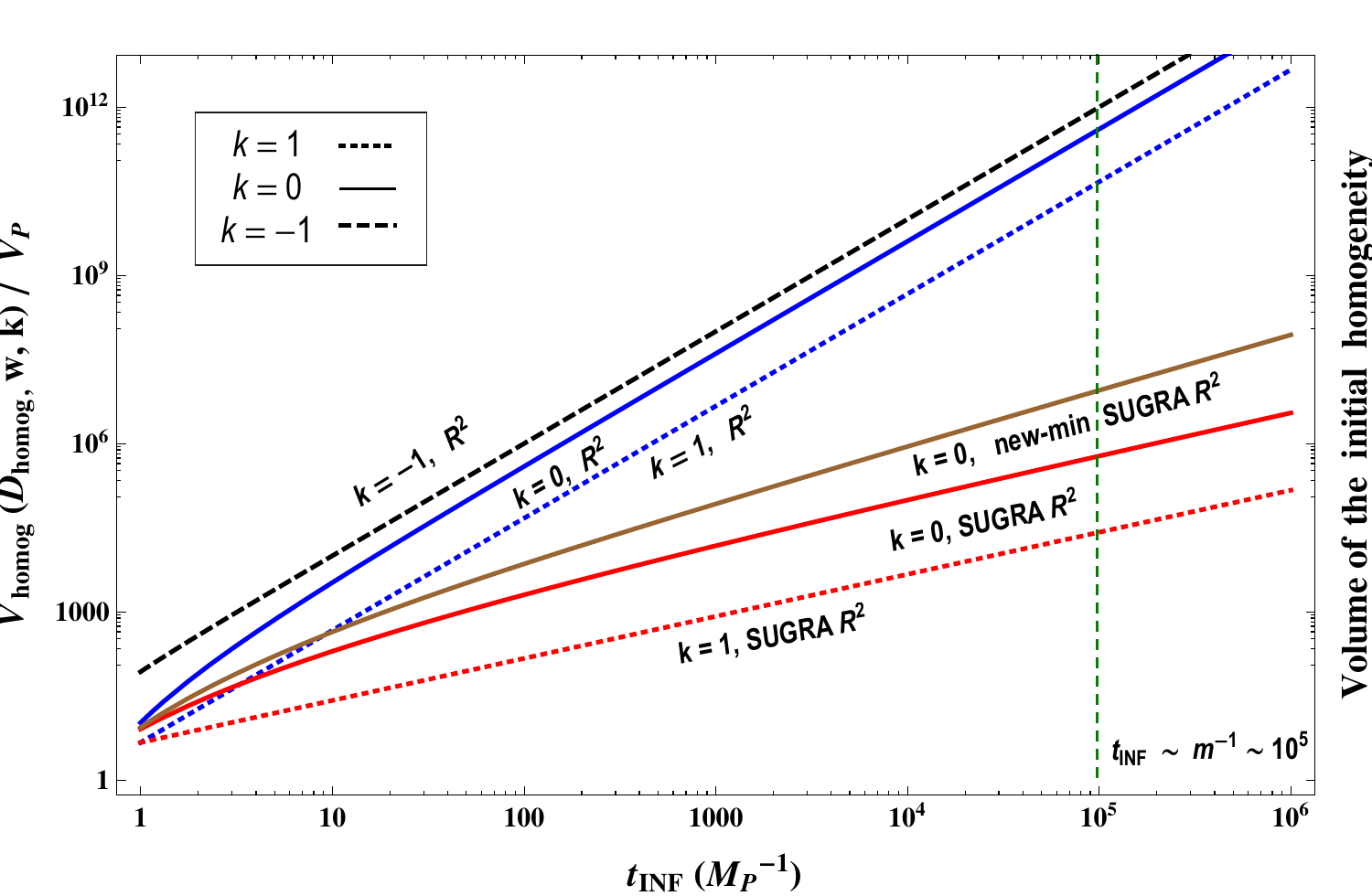} 
\caption {\small {The figure depicts the number of the causally disconnected regions (\# CDR) right after the Planck time required to be homogeneous in order for inflation to start and it manifests the initial conditions problem for the low-scale inflationary models in general.
The horizontal axis is the time that inflation starts. For the $R+R^2$ (super)gravity models the $t_\text{INF}$ is fixed by the CMB with $t_\text{INF}\sim m^{-1}\sim 10^5\, t_P$. 
}}

\end{figure}

\section{Conclusions}

We investigated the plateau inflationary potentials and focused in particular on $R+R^2$ gravity and supergravity models for inflation -  known also as Starobinsky models.  Plateau potentials are particularly motivated after the release of the P{\scshape lanck} 2013 results, however,  they account for low energy scale  inflaton models, requiring a rather extended acausal homogeneity in order for inflation to occur. We demonstrated that the problematic issue  of the initial conditions is less severe if supergravity is realized in nature due to the extra directions in the field space that can implement a relatively fast expansion rate before inflation. For flat  (closed) background geometry for the Universe, the $R+R^2$ gravity requires a huge initial homogeneous patch (huge initial 3-sphere) that contains about $10^{11}$ causally disconnected sub-patches while in the $R+R^2$ supergravity this number is  about $10^5-10^6$ times smaller, see figure 4.

We considered topologically trivial FLRW geometries.  The homogeneous patch of radius $D_\text{homog}$  is enclosed in a smaller volume when $k=1$ and in a larger one when $k=-1$:
\be \label{hier}
\# \,\text{CDR}(\text{closed}) < \# \,\text{CDR} (\text{flat}) < \# \,\text{CDR}(\text{open})\,.
\ee

In addition, the study of the pre-inflation supergravitational dynamics revealed interesting features such as the initial conditions that give sufficient number of e-foldings, that can  avoid the eternal process of self-reproduction, and generate a remarkable, however ephemeral, anisotropy \cite{Dalianis:2015fpa}.  
The investigation of the initial conditions for inflation can give us insights about the UV completion of gravity and might indicate a non-trivial topology for the Universe.

Our results, regarding the initial conditions problem and for a trivial topology, point towards $k=1$ and theories with extra field dimensions such as the $R+R^2$ supergravity theory.

\section*{Acknowledgments}

This talk was based on a work carried out with Fotis Farakos, to whom I am thankful for our collaboration,  his comments and our numerous discussions.

\end{document}